\documentclass{rectifyLDE-class}
\pdfoutput=1

\title{\Large
\textsc{A Conceptual Shift to Rectify a	Defect}			\\
\textsc{in the Lorentz--Dirac Equation}
}

\author{M. A. Oliver
\thanks{Previously: Department of Mathematical Sciences,
University of Bath, UK}						\\ [1.0ex]
\normalsize\textit{303 Glenalmond Avenue,}			\\
\normalsize\textit{Cambridge, CB2 8DT, UK}			\\ [1.0ex]
\normalsize{email: \texttt{MAOliver137@gmail.com}}		\\
}

\date{\normalsize\today}

\begin{document}

\maketitle

\begin{abstract}
In his analysis of the Classical Theory of Radiating Electrons, Dirac
(1938) draws attention to the characteristic instability of solutions
to the third order equation of motion.
He remarks that changing the sign of the self-force eliminates the
runaway solutions and gives `reasonable behaviour'.
Dirac rejects such a change and proceeds with an ad hoc modification
to the solutions of the initial value problem that is not consistent
with the principle of causality.
We argue that his reasons for rejecting the change of sign are
invalid on both physical and mathematical grounds.

The conceptual shift is to treat the physical particle as a composite
of the source particle and the energy-momentum that is reversibly
generated in its self-field by its motion.
The reversibly generated energy in the self-field is interpreted as
kinetic energy, and the changes that follow result in Dirac's change
of sign.
Several exact solutions to the new equation of motion and its linearisation
are given.
For a particle in orbital motion the self-force enables the applied
force to generate radiation and kinetic energy in the self-field
that results in an outward spiral motion.
The theory is consistent with all well-established principles of physics,
including the principle of causality.
\end{abstract}
\vfill

\pagebreak


\section{Introduction}

The system consists of a point particle with an applied force,
the particle having the physical attributes of mass~$m$ and
charge~$\elec$ with its field being defined by the Liénard
retarded potential.
The energy-momentum that is generated in its self-field by the
acceleration of the source particle causes a very small perturbation
to the motion, and we have to find the corresponding force.


\subsection{The Lorentz--Dirac Equation}

Solving the Lorentz--Dirac equation of motion as an initial value
problem results in unstable solutions in which the very
small perturbation of the self-force gives rise to an enormous
effect on the motion.
\citet[p.157]{Dirac:1938} remarked that this runaway behaviour indicates
an error in the sign of the self-force; he points out that changing the
sign results in solutions giving `reasonable behaviour'.
However, he rejected the change because it makes a particle in a Coulomb
field `spiral outwards, instead of spiralling inwards~\dots~as it
should in the classical theory'.

Dirac does not explain why the particle should spiral inwards in the
classical theory.
Of course, if it is assumed that the radiation reaction can be
modelled by a frictional force proportional to the velocity then the
solution of the resulting second order equation of motion does indeed
predict the particle spiralling inwards.
But the radiation reaction is, without doubt, proportional
to the hyper-acceleration, and the behaviour of a singular third order
differential equation cannot be inferred from one of second order;
thus, the objection is invalid.

In order to eliminate runaway behaviour, Dirac proposed an ad hoc
method of solution that results in acausal behaviour, manifested as
pre-acceleration.
\citet{Rohrlich:Particles} gives a detailed development.
However, these are not solutions of the initial value problem,
and they exhibit acausal behaviour; they are, therefore,
unacceptable.


\subsection{The Physics of Classical Electrodynamics}

The energy-momentum generated in the self-field of a charged particle
by its motion consists of two distinct parts.
The part irreversibly generated is the radiated energy-momentum; it
is propagated from the particle world-line as it is generated thereby
gaining its independence from its source particle.
Whereas, as \citet{Teitelboim:1970} has shown in detail, the reversibly
generated part is bound to its source particle.
Our conceptual shift is to treat the physical particle as a composite
of the source particle together with that part of the energy-momentum
that is reversibly generated in its self-field, which, therefore,
contributes to the total kinetic energy of the physical particle.
Clearly, this has some similarity to the self-energy in quantum
electrodynamics, though different in detail.

Energy-momentum generated in an applied field by the motion of a
particle does \emph{not} contribute to the self-field kinetic
energy of the physical particle.
Thus, there is a clear distinction between the process that generates
energy-momentum in the self-field and the process that generates
energy-momentum in an applied field:
the two physical processes are different.

Two postulates---constitutive assumptions---are required in order to
relate these energy-momentum flows to the corresponding forces
acting on their generating particle.
The force due to an applied field is postulated to be the familiar
Lorentz force.
The force due to the self-field is postulated to be
proportional to and opposite to the Abraham four-vector.\footnote{The
sign of this self-force is the opposite of that in the Lorentz--Dirac
equation.}
The choice of sign ensures that the reversible energy-momentum generated
by and bound to the source particle in the form of kinetic energy is
\emph{added} to the inertial kinetic energy.
Thus, within the conventional methodological framework, the theory
is modified and we get Dirac's change of sign.


\subsection{The Self-Force}

The self-force is the highest order (third) derivative in the equation
of motion and is, in general, a very small perturbation on the other
forces, therefore, the equation of motion is singular in the sense of
singular perturbation theory \citep{Hinch}.
It is characteristic of such singular equations that the sign of
the highest derivative determines whether the perturbation
term has a small effect on the motion or a destabilising effect leading
to `runaway' behaviour.

Another characteristic is the existence of two time scales.
During a brief initial period, the initial motion is exponentially
damped---this is justly termed radiation damping.
In the longer time scale the motion is dominated by the applied force
with the self-force perturbing the motion by generating energy-momentum
in the self-field.
The self-force interacts with the applied force, extracting
energy-momentum that is radiated and contributes to the
total kinetic energy of the physical particle.
Evidently, the self-force is not a simple `resistive' force.

Solutions of the new equation of motion, solved as an initial value
problem, have `reasonable behaviour'.
Their physical interpretation is straightforward, in terms of the
balance of energy-momentum flows.
The theory is mathematically well-defined, self-consistent, and
conforms to all well-established principles, including the principle
of causality.


\subsection{Structure of the Paper}

In Section~\ref{section: Electrodynamic Field}, we review the physical
interpretation of how the electrodynamic field depends on the motion
of its source particles.
The field, the charge current density, and the energy-momentum density
tensor are, ultimately, defined in terms of the Liénard potential.
Lorentz invariant identities for pairwise interactions and
self-interactions are established for the energy-momentum density
tensor.
In Section~\ref{section: Particle-Field Interactions}, we are concerned
with the action of the electrodynamic field on its source particle and
on other particles.
This requires two postulates: one for the relation between a particle
and the applied field---the Lorentz force; and the other for the relation
between a source particle and its self-field---the self-force.
Hence, the equation of motion.
Finally, in Section~\ref{section: Applications}, the initial value
problem for the equation of motion is solved for some important
special cases and their physical interpretations given.


\subsection{Notation}

The speed of propagation of the electrodynamic field
is~$c$.
We define constants $b=\twothirds\elec^2/c^3$, $\tcl=b/m$,
and $\fcl=1/\tcl$.
The particle position is given by $\V{r}=(x,y,z)$, the velocity
$\V{v}=\V{\Dot{r}}$, acceleration ${\V{\Dot{v}}}$, and
hyper-acceleration ${\V{\Ddot{v}}}$,
where an overdot denotes the derivative with respect to time~$t$;
for a bracketed expression the dot is a superscript on the
closing bracket, for example, $\updot{\V{\Dot{v}}\bSP\V{v}}$.

The metric tensor, $\eta$, has signature $(-,-,-,+)$.
The scalar product is usually written as a dot-product.
Greek indices range from 1~to~4, and the summation convention
applies.
A point $x^\lambda$ has Euclidean coordinates $(x,y,z,ct)$,
or $(\V{r},ct)$, and the invariant world-line parameter is
defined by $\intD s=\gamma^{-1}\intD t$, where
$\gamma=(1-\V{v}\bSP\V{v}/c^2)^{-\frac{1}{2}}$.
The particle position is given by the function $X$;
the velocity by $v=\intD X/\intD s$, with components
$v^\lambda=(\gamma\V{v},\gamma c)$, acceleration by $a=\intD v/\intD s$,
and hyper-acceleration by $h=\intD a/\intD s$.
From these, $v\SP v=c^2$ and $a\SP a\leqslant 0$.
The gradient operator has components
$\GRAD_\alpha=\bigl(\GRAD_x,\GRAD_y,\GRAD_z,\GRAD_{ct}\bigr)$.


\section{Electrodynamic Field}
\label{section: Electrodynamic Field}

This section is concerned with how the electrodynamic field---defined
in terms of the Liénard retarded potential---depends on the motion of
its source particles.


\subsection{Preliminaries}

Let the particle world-line be given by $x=X(s)$, with $-\infty<s<+\infty$.
From an arbitrary field point $x$ we draw the \emph{backward}
null cone to intersect the particle world-line at a point denoted
by~$X(s_R)$.
The null-vector from this point on the particle world-line to the
field point is $R=x-X(s_R)$ so that
\[
R\SP R
=
\bigl(x-X(s_R)\bigr)
\SP
\bigl(x-X(s_R)\bigr)
=
0
\]
Given the particle world-line, the value of the world-line parameter
$s_R$ depends on the choice of the field point~$x$, that is,
$s_R=s_R(x)$.
The point $X(s_R)$ and value $s_R$ thus obtained are said to be
\emph{retarded} relative to~$x$.


\subsubsection{Liénard Potential and its Field}

We start with the definition of the \citet{Lienard:1898b} potential,
\[
\Phi(x)=\frac{\elec}{R\cdot v}v
\]
where $v=\Dtl[X]{s}(s_R)$.
Following directly from its definition,
\begin{equation}
\DIV{\Phi}=0
\quad\text{and}\quad
\Box\Phi=\frac{4\pi}{c}j
\label{eq: Phi identities}
\end{equation}
where the charge current density is defined by
\begin{equation}
j(x)
=
\elec c\integ{s}{-\infty}{\infty}\delta^{(4)}\bigl(x-X(s)\bigr)v(s)
\label{eq: j}
\end{equation}
Hence, from \eqref{eq: Phi identities}, $\DIV{j}=0$.

The tensor field that is associated with a particle is defined by
\[
F^{\mu\nu}=\GRAD^\mu\Phi^\nu-\GRAD^\nu\Phi^\mu
\]
in terms of which the Maxwell--Lorentz identities are
\begin{equation}
\GRAD_\lambda F_{\mu\nu}
+
\GRAD_\nu F_{\lambda\mu}
+
\GRAD_\mu F_{\nu\lambda}
=
0
~,\qquad
\GRAD_\mu F^{\mu\nu}
=
\frac{4\pi}{c}j^\nu
\label{eq: M-L}
\end{equation}


\subsubsection{Energy-Momentum Density Tensor}

In a system of $\mathscr{N}$ particles, energy-momentum is generated
by the motion of the particles in two distinct ways.
Firstly, the motion of one particle in the field of another particle
generates energy-momentum in the field through their pairwise interaction.
Secondly, the motion of an individual particle generates energy-momentum
in its self-field.

We consider a system of $\mathscr{N}$ particles, with the particles
identified by upper case Latin subscripts.
The total potential is the sum of the individual Liénard potentials,
and similarly their corresponding currents and fields:
\begin{equation}
\Phi_\Sscript{total}=\sum_{A=1}^\mathscr{N}\Phi_A
~,\qquad
j_\Sscript{total}=\sum_{A=1}^\mathscr{N} j_A
~,\qquad
F_\Sscript{total}=\sum_{A=1}^\mathscr{N} F_A
\label{eq: sums}
\end{equation}
Clearly, the identities \eqref{eq: Phi identities}, and \eqref{eq: M-L}
are satisfied by $\Phi_\Sscript{total}$, $j_\Sscript{total}$, and
$F_\Sscript{total}$.

The energy-momentum density tensor is defined as a function of two fields,
\[
\Theta(F_A,F_B)
=
\frac{1}{4\pi}
\Bigl(
  F_A\SP F_B
  -
  \quarter\eta\,\trace\bigl(F_A\SP F_B\bigr)
\Bigr)
\]
On noting the bilinear structure of $\Theta$,
and writing $\Theta_{AB}$ for $\Theta(F_A,F_B)$, we get
\[
\Theta_\Sscript{total}
=
\Theta(F_\Sscript{total},F_\Sscript{total})
=
\sum_{A=1}^\mathscr{N}\sum_{B=1}^\mathscr{N}
  \Theta_{AB}
\]

Since the $\mathscr{N}$ fields $F_A$, and therefore $F_\Sscript{total}$,
satisfy the Maxwell--Lorentz identities, we have the identities
\begin{equation}
\DIV\Theta_\Sscript{total}
=
\frac{1}{c}j_\Sscript{total}\SP F_\Sscript{total}
=
\frac{1}{c}
\sum_{A=1}^\mathscr{N}\sum_{B=1}^\mathscr{N} j_A\SP F_B
\label{eq: div Theta=jF total}
\end{equation}
and
\begin{equation}
\DIV\Theta_{AA}
=
\frac{1}{c}j_A\SP F_A
~,\qquad A=1,\dotsc,\mathscr{N}
\label{eq: div Theta=jF}
\end{equation}
The pairwise interaction energy-momentum density tensor is defined by
\begin{equation}
\Theta_\Sscript{pairs}
=
\Theta_\Sscript{total}
-
\sum_{A=1}^\mathscr{N}\Theta_{AA}
\label{eq: Theta_AA=total-pair}
\end{equation}
Hence, with the aid of \eqref{eq: div Theta=jF total} and
\eqref{eq: div Theta=jF}, from \eqref{eq: Theta_AA=total-pair} we get
\begin{equation}
\DIV\Theta_\Sscript{pairs}
=
\frac{1}{c}
\sum_{A=1}^\mathscr{N}
  \sum^\mathscr{N}_{\begin{subarray}{c}B=1\\B\neq A\end{subarray}}j_A\SP F_B
\label{eq: div Theta=jF pair}
\end{equation}


\subsection{Physical Significance of the Field Interactions}

Both \eqref{eq: div Theta=jF} and \eqref{eq: div Theta=jF pair} are
interpreted as conservation laws.
Here, we examine physical significance of what is being conserved,
which is not always clear in the literature although the mathematical
development is well-known.


\subsubsection{$\Theta$ in Terms of $\V{E}$ and $\V{B}$}

In terms of the spatial field vectors $\V{E}=(E_x,E_y,E_z)$ and
$\V{B}=(B_x,B_y,B_z)$, the field components are
\[
F^{\lambda\nu}
=
\begin{bmatrix}
  0  & -B_z &  B_y & E_x	\\
 B_z &  0   & -B_x & E_y	\\
-B_y &  B_x &   0  & E_z	\\
-E_x & -E_y & -E_z &   0
\end{bmatrix}
\]
In terms of these variables, the energy-momentum density tensor has the form
\[
\Theta^{\lambda\mu}
=
\begin{bmatrix}
-\V{T} & \V{P}						\\
 \V{P} & U
\end{bmatrix}
=
\frac{1}{4\pi}
\begin{bmatrix}
\V{E}\V{E}+\V{B}\V{B}
-
\half\bigl(\V{E}\bSP\V{E}+\V{B}\bSP\V{B}\bigr)\V{I}
&
\V{E}\btimes\V{B}					\\
\V{E}\btimes\V{B}
&
\half
\bigl(
  \V{E}\bSP\V{E}
  +
  \V{B}\bSP\V{B}
\bigr)
\end{bmatrix}
\]
where $U=U(\V{E},\V{B})$, $\V{P}=\V{P}(\V{E},\V{B})$, and
$\V{T}=\V{T}(\V{E},\V{B})$ are the scalar energy density, the
Poynting vector, and the Maxwell stress tensor, respectively.
In terms of $\V{E}$ and $\V{B}$ the identities \eqref{eq: div Theta=jF} and
\eqref{eq: div Theta=jF pair} yield differential identities with the form
\begin{equation}
\label{eq: differential identities}
\begin{aligned}
\frac{1}{c}\Dpl[\V{P}]{t}
-
\Div{\V{T}}
&=
-
\Bigl(\rho\V{E}+\frac{1}{c}\V{j}\btimes\V{B}\Bigr)			\\
\frac{1}{c}\Dpl[U]{t}
+
\Div\V{P}
&=
-
\frac{1}{c}
\V{j}\bSP\V{E}
\end{aligned}
\end{equation}
The terms on the left-hand sides have a well-known interpretation:
they give the rates of change of field momentum and energy densities
at a space-time point, plus the influx of field momentum and energy
densities to the point.
Evidently, from \eqref{eq: j}, the right-hand sides are zero except
on particle world-lines.

Noting \eqref{eq: j}, the four-momentum is obtained from the
four-momentum density by spatial integration, and similarly
for other quantities.
From \eqref{eq: sums},
\[
\displaystyle{\V{E}_\Sscript{total}=\sum_{A=1}^\mathscr{N}\V{E}_A}
~,\qquad
\displaystyle{\V{B}_\Sscript{total}=\sum_{A=1}^\mathscr{N}\V{B}_A}
\]
and comparing \eqref{eq: differential identities} with
\eqref{eq: div Theta=jF total} we see that, for all $A$ and $B$,
\begin{equation}
\frac{\elec}{c}v_A\SP F_B
\quad\text{and}\quad
-\elec
\biggl(
  \V{E}_B+\frac{1}{c}\V{v}_A\btimes\V{B}_B
  \,,\;
  \V{v}_A\bSP\V{E}_B
\biggr)
\label{eq: identification}
\end{equation}
are equivalent, the field being evaluated on the world-line of its
source particle.
These expressions give the flow of energy-momentum generated in the
field labelled~$B$ by the motion of the particle labelled~$A$.


\subsubsection{Energy-Momentum of the Self-Field}

The self-field is propagated from its source particle on the
forward light-cone, its value at any field point depending on
the state of the motion of its source particle at the apex of
the light-cone.
From \eqref{eq: identification}---see also \eqref{eq: div Theta=jF}---the
rate at which the energy-momentum of the self-field of a
source particle is generated by the motion of the source particle is
$(\elec/c)v_A\SP F_A$.
After a non-trivial calculation based on the Liénard potential,
this quantity is given by the well-known expression:
\begin{equation}
\frac{\elec}{c}v_A\SP F_A
=
-bK
\label{eq: -bK}
\end{equation}
where $b=\twothirds\elec^2/c^3$ and
\begin{equation}
K
=
\Bigl(\eta-\frac{1}{c^2}vv\Bigr)\SP h
=
h+\frac{1}{c^2}a\SP a\,v
\label{eq: K}
\end{equation}
\citet{Rowe:1978} derived \eqref{eq: -bK} using distribution theory;
clear derivations are given by \citet{Teitelboim:1971} and
\citet{Hogan:1973a}, albeit with renormalisation; \citet{Dirac:1938}
involved both the retarded and advanced potentials together with
renormalisation.

The contravariant components of $K$ are given by
$\displaystyle{K^\lambda=\Bigl(\gamma\V{K},\frac{1}{c}\gamma\V{K}\bSP\V{v}\Bigr)}$,
with
\begin{equation}
\V{K}
=
\gamma^2\V{\Ddot{v}}
+
\frac{\gamma^4}{c^2}\V{\Ddot{v}}\bSP\V{v}\V{v}
+
3\frac{\gamma^4}{c^2}\V{\Dot{v}}\bSP\V{v}\V{\Dot{v}}
+
3\frac{\gamma^6}{c^4}(\V{\Dot{v}}\bSP\V{v})^2\V{v}
\label{eq: vector K}
\end{equation}
first obtained by \citet{Abraham:1903} within a classical kinematical
framework.
With the aid of the identity
$\V{\Dot\gamma}=\gamma^3\V{\Dot{v}}\bSP\V{v}/c^2$, we find
\begin{equation}
\V{K}\bSP\V{v}
=
\gamma^4\V{\Ddot{v}}\bSP\V{v}
+
3\frac{\gamma^6}{c^2}(\V{\Dot{v}}\bSP\V{v})^2
\label{eq: K.v}
\end{equation}

On rearranging \eqref{eq: K.v}, with the aid of \eqref{eq: -bK},
we can write the temporal component of $(\elec/c)v_A\SP F_A$ as
\begin{equation}
-b\V{K}\bSP\V{v}
=
S^\Sscript{(rad)}
-
S^\Sscript{(rev)}
\label{eq: S-def}
\end{equation}
where
\begin{equation}
S^\Sscript{(rad)}
=
b\gamma^4\Bigl(\V{\Dot{v}}\bSP\V{\Dot{v}}
+
\frac{\gamma^2}{c^2}(\V{\Dot{v}}\bSP\V{v})^2\Bigr)
=
-b\,a\cdot a
\geqslant
0
\label{eq: S-rad}
\end{equation}
is the radiated power, first obtained by \citet{Lienard:1898b},
irreversibly generated in the self-field by the motion of its
source particle.
And
\begin{equation}
S^\Sscript{(rev)}
=
b\updot{\gamma^4\V{\Dot{v}}\bSP\V{v}}
\label{eq: S-kinetic}
\end{equation}
is the power reversibly generated in the self-field by the motion
of its source particle.


\section{Particle-Field Interactions}
\label{section: Particle-Field Interactions}

Following the introduction of our conceptual shift, the forces acting
on a particle are identified and, thereby, we obtain the equation of
motion for a charged particle.


\subsection{The `Dressed' Particle}

The reversibly generated energy-momentum in the self-field,
$S^\Sscript{(rev)}$, is bound to the source particle
\citep[Section~III]{Teitelboim:1970}.
Our conceptual shift is to treat the physical particle as a composite
of the source particle together with that part of the energy-momentum
that is reversibly generated in its self-field.
On integrating \eqref{eq: S-kinetic} we get
\[
T_\Sscript{charge}
=
b\gamma^4\V{\Dot{v}}\bSP\V{v}
\]
where $T_\Sscript{charge}$ is interpreted as the kinetic energy
generated in the self-field of the charged particle.
Accordingly, the \emph{total} kinetic energy of the physical particle
is defined as the sum of the inertial kinetic energy of the source
particle and the kinetic energy generated in its self-field:
\begin{equation}
T
=
T_\Sscript{mass}
+
T_\Sscript{charge}
=
mc^2(\gamma-1)+b\gamma^4\V{\Dot{v}}\bSP\V{v}
\label{eq: T}
\end{equation}
We remark that if $\V{\Dot{v}}\bSP\V{v}\geqslant0$ and
$\updot{\gamma^4\V{\Dot{v}}\bSP\V{v}}\geqslant0$ then the self-field
kinetic energy is increasing, and that the sign of $S^\Sscript{(rev)}$
in \eqref{eq: S-def} reflects the fact that it is bound to the particle,
whereas $S^\Sscript{(rad)}$ is propagated away from the particle.

It is tempting to say that the `bare' source particle is `dressed' with
the reversibly generated energy-momentum in its self-field.
However, any analogy to the self-energy in quantum electrodynamics
cannot to be taken too literally.


\subsection{Balance of Forces on a Particle}

The fundamental law of particle dynamics states that for each particle
the forces that act on it must balance.
In the classical electrodynamics of $\mathscr{N}$ particles there
are $\mathscr{N}$ equations of the form
\begin{equation}
f_\Sscript{charge}
+
f_\Sscript{mass}
+
f_\Sscript{Lorentz}
+
f_\Sscript{applied}
=0
\label{eq: balance-of-forces}
\end{equation}
where $f_\Sscript{charge}$ is the four-force on the particle due to the
changes that its motion induces on its self-field, $f_\Sscript{mass}$
is the inertial four-force due to the particle mass,
$f_\Sscript{Lorentz}$ is the four-force on the particle due to
pairwise interactions with the other $\mathscr{N}-1$ particles,
and $f_\Sscript{applied}$ is an externally applied four-force.


\subsection{Inertial Force}

The constitutive law for the inertial force:
\begin{postulate}[Inertial Force]
The inertial force acting on a particle of mass $m$ is
\begin{equation}
f_\Sscript{mass}
=
-ma
\label{eq: f-mass}
\end{equation}
where $a$ is the four-acceleration of the particle.
\end{postulate}
Concerning the concept of inertial force see \citet{Noll:2007}.


\subsection{Electrodynamic Forces}

In \eqref{eq: identification} we have the expression $(\elec/c)v_A\SP F_B$
for the flow of energy-momentum generated in the field labelled~$B$ by the
motion of the particle labelled~$A$---the expression being valid for all
$A$~and~$B$.
There are two physical processes to consider:
\begin{itemize}
\item
For $B\neq A$ we have a pairwise field interaction in which
the energy-momentum $(\elec/c)v_A\SP F_B$ is generated in
the field of another particle.
\item
For $B=A$ we have the self-field interaction in which
the energy-momentum $(\elec/c)v_A\SP F_A$ is generated
in the field of the particle.
As it is generated, part of this self-field energy-momentum separates
from its source as radiation, with the remaining part being bound to
its source particle thereby contributing to the total kinetic energy
of the physical particle---see \eqref{eq: -bK} and \eqref{eq: S-def}.
\end{itemize}
Manifestly, the physical interaction between a particle and its
self-field differs radically from the physical interaction between
a particle and the field of another particle.

We now need to relate the rate at which energy-momentum is generated
by the motion of the particle to the resulting forces acting on the
particle.
Since we have two distinct physical processes it is evident that each
has to be dealt with separately: one for the pairwise field interaction
and the other for the self-field interaction.


\subsubsection{Lorentz Force}

For the pairwise field interaction: we make the assumption that
the flow of energy-momentum that is generated in the field $F_B$
by the motion of particle~$A$ is balanced by a four-force $f_A$
acting on the particle~$A$.
\[
f_A+\frac{\elec}{c}v_A\SP F_B=0
\]
Hence,
\begin{postulate}[Lorentz Force]
The four-force exerted on a particle at the point $x$ by the field
of another particle is given by
\begin{equation}
f_\Sscript{Lorentz}(F)=-\frac{\elec}{c}v\SP F(x)
\label{eq: f-Lorentz}
\end{equation}
where $\elec$ is the particle charge, $v$ its four-velocity, and $F$
the field of another particle.
\end{postulate}


\subsubsection{Self-Force}

For the self-field interaction, we make the physical assumption that
the energy-momentum generated in the self-field by the motion of its
source particle is, \emph{at the point at which it is generated},
part of the particle--self-field composite---the radiation then being
propagated away from its point of generation.
Accordingly, the flow of inertial energy-momentum \emph{to} the
particle is $-ma$, giving a four-force of~$f_\Sscript{mass}$ on
the particle.
And from \eqref{eq: identification}, the energy-momentum generated
\emph{in}---that is, flowing \emph{to}---the self-field
is~$(\elec/c)v_A\SP F_A$, giving a four-force
of~$f_\Sscript{charge}$ on the particle.
Thus, noting \eqref{eq: -bK}, the mass and the charge give rise
to the four-force
\begin{equation}
f_\Sscript{mass}+f_\Sscript{charge}=-ma-bK
\label{eq: shift}
\end{equation}
acting on the particle.
Thus,
\begin{postulate}[Self-Force]
The four-force that the self-field, $F$, exerts on its source
particle is given by
\begin{equation}
f_\Sscript{charge}=\frac{\elec}{c}v\SP F
\label{eq: f-charge}
\end{equation}
where $\elec$ is the particle charge and $v$ its four-velocity,
the product $v\SP F$ being evaluated on the world-line.
Without approximation, we have
\begin{equation}
f_\Sscript{charge}=-bK
\label{eq: f-charge=-bK}
\end{equation}
where $K$ is the Abraham four-vector and $b=\twothirds\elec^2/c^3$.
\end{postulate}

The contrast between \eqref{eq: f-Lorentz} and \eqref{eq: f-charge}
is evident: in the former, the flow of energy-momentum is to the applied
field, whereas, in the latter, the energy-momentum is generated as a part
of the particle--self-field composite.


\subsection{Equation of Motion}

The exact equation of motion for a single particle is obtained by
setting $f_\Sscript{Lorentz}=0$ in \eqref{eq: balance-of-forces} and
substituting from \eqref{eq: f-mass} and \eqref{eq: f-charge=-bK}:
\begin{equation}
bK+ma=f_\Sscript{applied}
\label{eq: equation of motion}
\end{equation}

The spatial component is
\begin{equation}
b\V{K}
+
m\updot{\gamma\V{v}}
=
\V{f}_\Sscript{applied}
\label{eq: spatial equation of motion}
\end{equation}
and, on substituting from \eqref{eq: S-def}, using \eqref{eq: S-kinetic},
and rearranging, the temporal component is
\begin{equation}
mc^2\V{\Dot{\mathnormal{\gamma}}}
+
b\updot{\gamma^4\V{\Dot{v}}\bSP\V{v}}
-
S^\Sscript{(rad)}
=
P_\Sscript{applied}
\label{eq: spatial equation of power}
\end{equation}
where $P_\Sscript{applied}=\V{f}_\Sscript{applied}\bSP\V{v}$ is the power
generated on the particle by the applied force.
With the definition of the total kinetic energy given in
\eqref{eq: T}, the balance of power,
\eqref{eq: spatial equation of power}, is
\begin{equation}
\V{\Dot{\mathnormal{T}}}
=
S^\Sscript{(rad)}
+
P_\Sscript{applied}
\label{eq: power balance}
\end{equation}
If we interpret radiation as a flow of heat then \eqref{eq: power balance}
is, for one source particle and its self-field, an expression of the first
law of thermodynamics.


\section{Applications}
\label{section: Applications}

The solutions to the applications reveal that the underlying physical
explanation for the behaviour of the particle is due to the self-force
extracting energy-momentum from the applied force and converting it to
radiation and energy-momentum for the self-field of the particle.
The applied forces considered here are constant and, therefore, provide
an unlimited source of energy-momentum.
For a free particle, the only energy-momentum available for radiation
is from the initial acceleration.
In all cases the self-force generates only a very small effect, which
is consistent with its very small magnitude.

If no exact solution to the equation of motion can be found then
we are reduced to singular perturbation methods: an accessible
introductory account of the methods of Matched Asymptotic Expansions
and of Multiple Scales is to be found in \citet{Hinch}.
However, for motion that is not too extreme, the equation of motion
can be approximated by a linearised equation of motion that can
be solved without further approximation in several important special
cases.

We remark that the corresponding solutions to the Lorentz--Dirac
equation, solved as an initial value problem, are obtained by setting
$b\rightarrow-b$, which implies that we set $\tcl\rightarrow-\tcl$
and $\fcl\rightarrow-\fcl$.
On recalling that, for an electron, the interval
$\tcl=\twothirds\elec^2/mc^3\approx6\times10^{-24}$ seconds,
the unstable behaviour of such solutions is manifest.


\subsection{The Exact Equation of Motion}

The integration of the exact equation of motion for a free particle,
a particle given an applied impulse, and for a particle with a
frictional four-force is straightforward.


\subsubsection{Free Particle}

From \eqref{eq: equation of motion}, recalling \eqref{eq: K}, the equation
of motion for a free particle is
\begin{equation}
h+\frac{1}{c^2}a\SP av+\fcl a=0
\label{eq: free}
\end{equation}
to be integrated with initial values $v(0)=v_0$ and $a(0)=a_0$.

Taking the scalar product of \eqref{eq: free} with $a$, we get
\[
\Dtl{s}(a\cdot a)+2\fcl a\cdot a=0
\]
Integrating yields
\begin{equation}
a\cdot a=a_0\cdot a_0\,\E^{-2\fcl s}
\label{eq: S rad}
\end{equation}
Using this to eliminate the factor $a\cdot a$ in
\eqref{eq: free} yields the linear equation
\[
\DDtl[v]{s}
+
\fcl\Dtl[v]{s}
-
\sigma^2\fcl^2\E^{-2\fcl s}v=0
~,\qquad
\sigma=\tcl\sqrt{-a_0\cdot a_0/c^2}
\]
With a change of variable, setting $x=\E^{-\fcl s}$ and $v(s)=w(x)$,
we find
\[
\DDtl[w]{x}-\sigma^2w=0
\]
Hence,
\begin{equation}
v=v_0\cosh\bigl[\sigma(1-\E^{-\fcl s})\bigr]+
  a_0\frac{\tcl}{\sigma}\sinh\bigl[\sigma(1-\E^{-\fcl s})\bigr]
\label{eq: free solution}
\end{equation}
For $s\gg\tcl$ the velocity has the constant value
$\displaystyle{v_0\cosh\sigma+\tcl a_0\frac{\sinh\sigma}{\sigma}}$.

Substituting \eqref{eq: S rad} into \eqref{eq: S-rad}, we get
$S^\Sscript{(rad)}=-ba_0\SP a_0\E^{-2\fcl s}$,
and with \eqref{eq: power balance},
\[
\V{\Dot{\mathnormal{T}}}
=
S^\Sscript{(rad)}
\]
Thus, the self-field kinetic energy from the initial acceleration is
converted into radiated energy.
This phenomenon can rightly be described as radiation damping.
From $s=0$ to $s=\infty$, the total energy radiated is
$T(0)-T(\infty)=-\half\tcl ba_0\SP a_0\geqslant0$.


\subsubsection{Applied Impulse}

The equation of motion for a free particle with an impulse
$f_\Sscript{applied}=k\delta(s-s_i)$ is
\begin{equation*}
h+\frac{1}{c^2}a\SP av+\fcl a=k\delta(s-s_i)
\end{equation*}
to be integrated with initial values $v(0)=v_0$ and $a(0)=a_0$.
Noting the solution for a free particle, \eqref{eq: free solution}, we find
\begin{multline*}
v=v_0\cosh\bigl[\sigma(1-\E^{-\fcl s})\bigr]+
  a_0\frac{\tcl}{\sigma}\sinh\bigl[\sigma(1-\E^{-\fcl s})\bigr]
\\
+
k\frac{\tcl}{\sigma}\sinh\bigl[\sigma(1-\E^{-\fcl(s-s_i)})\bigr]
\Heaviside(s-s_i)
\end{multline*}
where $\Heaviside$ is the Heaviside step function and
$\sigma=\tcl\sqrt{-a_0\cdot a_0/c^2}$.


\subsubsection{Frictional Force}

From \eqref{eq: K} and \eqref{eq: equation of motion}, the equation
of motion for a particle with the frictional four-force
$f_\Sscript{applied}=-kv$, where $k$ is a constant, is
\begin{equation}
h+\frac{1}{c^2}a\SP av+\fcl a+\frac{k}{b}v=0
\label{eq: frictional}
\end{equation}
with initial values of $v_0$ and $a_0$ for the velocity and acceleration,
respectively.
Following the same procedure as for the free particle, using
\eqref{eq: S rad} to eliminate the factor $a\cdot a$ in \eqref{eq: frictional},
yields the linear equation
\[
\DDtl[v]{s}
+
\fcl\Dtl[v]{s}
+
\biggl(\frac{k}{b}-\sigma^2\fcl^2\E^{-2\fcl s}\biggr)v
=
0
~,\qquad
\sigma=\tcl\sqrt{-a_0\cdot a_0/c^2}
\]

With the substitution $v(s)=w\bigl(x(s)\bigr)\exp\bigl(-\half\fcl s\bigr)$
and $x=\sigma\E^{-\fcl s}$, we find
\begin{equation}
x^2
\DDtl[w]{x}
+
x
\Dtl[w]{x}
-
\bigl(
  x^2
  +
  \alpha^2
\bigr)w
=
0
\label{eq: Bessel}
\end{equation}
where $\alpha^2=\quarter\bigl(1-4\tcl k/m\bigr)$.
The modified Bessel equation, \eqref{eq: Bessel}, has the solution
\[
w(x)
=
C_1I_\alpha(x)+C_2K_\alpha(x)
\]
where $I_\alpha$ and $K_\alpha$ are modified Bessel functions of
order~$\alpha$.
Hence,
\[
v(s)
=
\bigl\{
  C_1I_\alpha(\sigma\E^{-\fcl s})
  +
  C_2K_\alpha(\sigma\E^{-\fcl s})
\bigr\}
\exp\bigl(-\half\fcl s\bigr)
\]
where, applying the initial conditions, the constants $C_1$ and $C_2$ are
given by
\begin{align*}
C_1
&=
\frac{1}{\Delta}
\biggl(
  \frac{1}{\sigma\fcl}
  K_\alpha(\sigma)
  \bigl(\V{a}_0+\half\fcl\V{v}_0\bigr)
  +
  K'_\alpha(\sigma)\V{v}_0
\biggr)
\\
C_2
&=
\frac{-1}{\Delta}
\biggl(
  \frac{1}{\sigma\fcl}
  I_\alpha(\sigma)
  \bigl(\V{a}_0+\half\fcl\V{v}_0\bigr)
  +
  I'_\alpha(\sigma)\V{v}_0
\biggr)
\end{align*}
with
\[
\Delta=I_\alpha(\sigma)K'_\alpha(\sigma)-K_\alpha(\sigma)I'_\alpha(\sigma)
\]


\subsection{The Linearised Equation of Motion}

For the linearised approximation only those terms that are linear
in $\V{v}$ and its derivatives are retained---for approximately
circular motion, this is equivalent to dropping the terms of order
$\V{v}\bSP\V{v}/c^2$.
From \eqref{eq: vector K} we find $\V{K}=\V{\Ddot{v}}$, and the equation
of motion \eqref{eq: spatial equation of motion} reduces to
\begin{equation}
b\V{\Ddot{v}}+m\V{\Dot{v}}=\V{f}_\Sscript{applied}
\label{eq: EofM-single-approx}
\end{equation}
to be integrated with initial conditions $\V{r}(0)=\V{r}_0$,
$\V{v}(0)=\V{v}_0$, and $\V{a}(0)=\V{a}_0$.
The radiated power, \eqref{eq: S-rad}, and the total kinetic energy,
\eqref{eq: T}, reduce to
\begin{equation}
S^\Sscript{(rad)}=b\V{\Dot{v}}\bSP\V{\Dot{v}}
\quad\text{and}\quad
T=\half m\V{v}\bSP\V{v}+b\V{\Dot{v}}\bSP\V{v}
\label{eq: S-rad-approx}
\end{equation}

For the first three applications, the source of the particle energy
is finite and is expended by the change in particle motion and
radiation.
By contrast, in the remaining four applications the particle is in
a field that is held constant, from which there is no limit to the
available power.
With the exception of the Coulomb field, the problems are solved
without further approximation.


\subsubsection{Free Particle}

From \eqref{eq: EofM-single-approx}, the equation of motion for a
free particle is
\[
b\V{\Ddot{v}}+m\V{\Dot{v}}=0
\]
with the solution
\begin{equation}
\V{v}=\V{v}_0+\tcl\bigl(1-\E^{-\fcl t}\bigr)\V{a}_0
\label{eq: free approx}
\end{equation}
Hence, from \eqref{eq: power balance} and \eqref{eq: S-rad-approx} we
find that
\[
\V{\Dot{\mathnormal{T}}}
=
S^\Sscript{(rad)}
=
b\V{a}_0\bSP\V{a}_0\E^{-2\fcl t}
\]
In the initial interval of $O(\tcl)$ the power contained in the
initial acceleration is dissipated as radiation, and for $t\gg\tcl$
the velocity is constant: $\V{v}\rightarrow\V{v}_0+\tcl\V{a}_0$.


\subsubsection{Applied Impulse}

With the applied force an impulse at $t=t_i>0$~, equation
\eqref{eq: EofM-single-approx} becomes
\[
b\V{\Ddot{v}}+m\V{\Dot{v}}=\delta(t-t_i)\tcl\V{k}
\]
where $\V{k}$ is constant.
The solution is
\[
\V{v}(t)
=
\V{v}_0+\tcl(1-\E^{-\fcl t})\V{a}_0
+
\tcl\bigl(1-\E^{-\fcl(t-t_i)}\bigr)((1/m)\V{k})\Heaviside(t-t_i)
\]
where $\Heaviside$ is the Heaviside step function.
The impulse gives the particle an additional acceleration of $(1/m)\V{k}$.
The final steady-state velocity is $\V{v}_0+\tcl\V{a}_0+(\tcl/m)\V{k}$~,
following the expenditure of the finite energy supplied by the initial
acceleration and the impulse.
The self-force both moderates the effects of the initial acceleration
and the impulse while generating radiation.


\subsubsection{Frictional Force}

Substituting the frictional force $\V{f}_\Sscript{applied}=-k\V{v}$ into
\eqref{eq: EofM-single-approx}, we get the equation of motion
\[
b\V{\Ddot{v}}+m\V{\Dot{v}}+k\V{v}=0
\]
Integrating,
\[
\V{v}(t)
=
\bigl\{
  \V{v}_0\cosh\alpha\fcl t
  +
  (1/\alpha\fcl)\bigl(\V{a}_0+\half\fcl\V{v}_0\bigr)\sinh\alpha\fcl t
\bigr\}
\exp\bigl(-\half\fcl t\bigr)
\]
where ${\alpha=\half\Bigl(1-\sqrt{1-4\tcl k/m}\,\Bigr)}$.
For any system of interest we have $\tcl k/m<1$.
The rate of decay of the initial velocity is increased slightly
by the self-force generating radiation.


\subsubsection{Constant Uniform Electric Field}

We have $\V{f}_\Sscript{applied}=\elec\V{E}$ with the electric
field $\V{E}$ constant.
The equation of motion \eqref{eq: EofM-single-approx} is
\[
b\V{\Ddot{v}}+m\V{\Dot{v}}=\elec\V{E}
\]
Integrating,
\[
\V{v}
=
\V{v}_0+\tcl\V{a}_0+(\elec/m)(t-\tcl)\V{E}
-
\tcl
\E^{-\fcl t}
\bigl(
\V{a}_0
-
(\elec/m)\V{E}
\bigr)
\]
Following the initial interval of $O(\tcl)$, the
acceleration is constant and the radiated power,
$S^\Sscript{(rad)}>0$.
Also, $S^\Sscript{(rev)}=S^\Sscript{(rad)}$, since
$\V{\Ddot{v}}=0$.
Thus, the particle is accelerated by the applied force and,
although there is no overall working by the self-force, the
self-force creates the radiation and the increasing kinetic
energy of the self-field.


\subsubsection{Harmonic Oscillator Field}

The force $\V{f}_\Sscript{applied}=-k\V{r}$, with $k>0$,
is the harmonic force centred at the origin.
With this applied force, the equation of motion \eqref{eq: EofM-single-approx}
is
\begin{equation}
b\V{\dddot{r}}+m\V{\Ddot{r}}+k\V{r}=0
\label{eq: harmonic motion}
\end{equation}
On taking the scalar product with $\V{r}\btimes\V{\Dot{r}}$, and noting that
$\V{r}\btimes\V{\Dot{r}}\bSP\V{\dddot{r}}
=\updot{\V{r}\btimes\V{\Dot{r}}\bSP\V{\Ddot{r}}}$,
we find
\begin{equation}
\Dtl{t}\bigl(
\V{r}\btimes\V{\Dot{r}}\bSP\V{\Ddot{r}}\bigr)
+
\fcl\V{r}\btimes\V{\Dot{r}}\bSP\V{\Ddot{r}}
=
0
\label{eq: settle equation}
\end{equation}
with the solution
\begin{equation}
\V{r}\btimes\V{\Dot{r}}\bSP\V{\Ddot{r}}
=
\V{r}_0\btimes\V{\Dot{r}}_0\bSP\V{a}_0\E^{-\fcl t}
\label{eq: settle solution}
\end{equation}
Therefore, any initial acceleration parallel to the angular velocity
decays, and the motion settles down to lie in the plane of the angular
velocity.

The equation of motion, \eqref{eq: harmonic motion}, on setting
$k=m\Omega^2$ and multiplying by $\tcl^3/b$ is
\[
\tcl^3\V{\dddot{r}}+\tcl^2\V{\Ddot{r}}+\lambda^2\V{r}=0
\]
where the parameter $\lambda=\tcl\Omega$---it is evident that
for any system of interest $\lambda\ll1$.

With the trial solution $\V{r}(t)=\E^{\alpha\fcl t}$, we find
$\alpha^3+\alpha^2+\lambda^2=0$.
This cubic has one real root and two complex roots that we write in the form
\[
\alpha_0=-1-2\kappa\tcl
~,\qquad
\alpha_1=\kappa\tcl-\I\omega\tcl
~,\qquad
\alpha_2=\kappa\tcl+\I\omega\tcl
\]
where the constants $\omega$ and $\kappa$ are given by
\begin{equation}
\omega\tcl
=
\textstyle\frac{\sqrt{3}}{2}(\Lambda^+-\Lambda^-)
~,\qquad
\kappa\tcl
=
\half(\Lambda^++\Lambda^-)-\third
\label{eq: omega and kappa}
\end{equation}
with
\[
\Lambda^\pm
=
\biggl[
\textstyle\frac{1}{27}+\half\lambda^2
\pm
\sqrt{\textstyle\frac{1}{27}\lambda^2+\quarter\lambda^4}\,
\biggr]^{1/3}
\]
and the constants satisfy the identity $\omega^2=3\kappa^2+2\fcl\kappa$.

The general solution of \eqref{eq: harmonic motion} can be
written in the form
\[
\V{r}(t)
=
\V{a}\E^{-(\fcl+2\kappa)t}
+
\E^{\kappa t}
\bigl[
  \V{b}\cos\omega t
  +
  \V{c}\sin\omega t
\bigr]
\]
where $\V{a}$, $\V{b}$, and $\V{c}$ are constants of integration.
On expanding $\Lambda^\pm$ in a power series, from \eqref{eq: omega and kappa},
we find
\[
\omega\tcl
=
\lambda+O\bigl(\lambda^3\bigr)
~,\qquad
\kappa\tcl
=
\half\lambda^2+O\bigl(\lambda^3\bigr)
\]
from which, to terms $O\bigl(\lambda^2\bigr)$, we find
$\omega=\Omega$ and $\kappa=\half\lambda\Omega$.

Following the initial interval of $O(\tcl)$, the motion takes place
in the plane defined by $\V{b}$ and~$\V{c}$ in a slowly expanding
elliptical logarithmic spiral with constant angular velocity.
The source of energy for this spiral motion, with the slowly increasing
speed of the particle and the radiation generated, is the interaction
of the self-force and the harmonic field.
Both the kinetic energy and the harmonic potential energy
increase with time.


\subsubsection{Constant Uniform Magnetic Field}

With an applied magnetic field the equation of motion
\eqref{eq: EofM-single-approx} is
\[
b\V{\Ddot{v}}+m\V{\Dot{v}}=\frac{\elec}{c}\V{v}\btimes\V{B}
\]
Let the magnetic field be $\V{B}=(0,0,B)$, with $B$ constant, and write
$\V{v}=(u,v,w)$ so that $\V{v}\btimes\V{B}=(Bv,-Bu,0)$.
Accordingly, the motion in the direction of the $z$-axis is that
of a free particle; from \eqref{eq: free approx},
\[
v_z=v_{z\,0}+\tcl\bigl(1-\E^{-\fcl t}\bigr)a_{z\,0}
\]

For the motion in the $xy$-plane we use matrix notation.
Let
\[
\mx{v}
=
\begin{bmatrix}
u \\
v
\end{bmatrix}
\quad\text{and}\quad
\mx{A}
=
\begin{bmatrix}
0  & 1 \\
-1 & 0
\end{bmatrix}
\]
then, with $\Omega=\elec B/mc$, the equation of motion, projected
onto the $xy$-plane, is
\begin{equation}
\tcl\V{\Ddot{\mx{v}}}+\V{\Dot{\mx{v}}}=\Omega\mx{A}\mx{v}
\label{eq: magn eqn of motion 2D}
\end{equation}

The solution to the equation of motion \eqref{eq: magn eqn of motion 2D} is
\begin{equation}
\mx{v}(t)
=
\E^{\kappa t}\E^{\mx{A}\omega t}\mx{h}
+
\E^{-(\fcl+\kappa)t}\E^{-\mx{A}\omega t}\mx{k}
\label{eq: solution-magn-field}
\end{equation}
where the column matrices $\mx{h}$ and $\mx{k}$ are to be determined by
the initial conditions,
\[
\mx{v}(0)=\mx{v}_0
~,\qquad
\V{\Dot{\mx{v}}}(0)=\mx{a}_0
\]
We remark that
\[
\E^{\mx{A}\omega t}
=
\begin{bmatrix} \cos\omega t & \sin\omega t \\
  -\sin\omega t &\cos\omega t\end{bmatrix}
\]
The two constants, both positive, are given by
\[
\omega
=
\half\fcl\sqrt{\half\sqrt{1+16\lambda^2}-\half}
~,\qquad
\kappa
=
\half\fcl\Bigl[\sqrt{1+4\tcl^2\omega^2}-1\Bigr]
\]
where $\lambda=\tcl\Omega$; they satisfy the identity
$\omega^2=\kappa^2+\fcl\kappa$.

It is evident that for any system of interest $\lambda\ll1$.
Expanding $\kappa$ and $\omega$ we find
\[
\omega=\Omega+O(\lambda^2)
~,\qquad
\kappa=\lambda\Omega+O(\lambda^3)
\]
With these approximations the solution \eqref{eq: solution-magn-field}
simplifies to
\[
\mx{v}(t)
=
\E^{\lambda\Omega t}\E^{\mx{A}\Omega t}\mx{h}
+
\E^{-\fcl t}\E^{-\mx{A}\Omega t}\mx{k}
\]

The motion of the particle, given in \eqref{eq: solution-magn-field},
projected onto the $xy$-plane, is in a slowly expanding logarithmic
spiral with constant angular velocity, modified in the initial interval
of~$O(\tcl)$.
Since $\mx{v}^\Trans\mx{A}\mx{v}=0$, there is no work
done by the magnetic field directly on the particle, and we have
$\V{\Dot{\mathnormal{T}}}=S^\Sscript{(rad)}$.
From \eqref{eq: solution-magn-field}, we find
\[
\V{\Dot{\mathnormal{T}}}
=
S^\Sscript{(rad)}
=
b\bigl(\kappa^2+\omega^2\bigr)\E^{2\kappa t}\mx{h}^\Trans\mx{h}
+
O\bigl(\E^{-\fcl t}\bigr)
\]
Manifestly, the source of energy for this spiral motion, with the
slowly increasing speed of the particle and the radiation generated,
is the magnetic field.


\subsubsection{Coulomb Field}

With the Coulomb force, the equation of motion \eqref{eq: EofM-single-approx}
becomes
\[
b\V{\dddot{r}}+m\V{\Ddot{r}}+\frac{\elec^2}{r^3}\V{r}=0
\]
On taking the scalar product with $\V{r}\btimes\V{\Dot{r}}$ we again
get the equation \eqref{eq: settle equation} with its solution
\eqref{eq: settle solution}.
Therefore, any initial acceleration parallel to the angular velocity
decays, and the motion settles down to lie in the plane through the
origin and perpendicular to the angular velocity.

Taking the scalar product of the equation of motion with $\V{\Dot{r}}$,
noting \eqref{eq: S-rad-approx} and rearranging, we get
\[
\Dtl{t}\biggl(T-\frac{\elec^2}{r}\biggr)
=
S^\Sscript{(rad)}
>0
\]
The total energy of the particle is slowly increasing: the source of this
energy being the power derived from the self-force.

In polar coordinates, the radial and transverse components of the
equation of motion are
\begin{align*}
\V{\Ddot{\mathnormal{r}}}-r\V{\Dot{\vartheta}}^2+\frac{\elec^2}{m}\frac{1}{r^2}
&=
-\tcl
\bigl(
  \V{\dddot{\mathnormal{r}}}-3\Dot r\V{\Dot{\vartheta}}^2
  -
  3r\V{\Dot{\vartheta}}\V{\Ddot{\vartheta}}
\bigr)									\\
2\Dot r\V{\Dot{\vartheta}}+r\V{\Ddot{\vartheta}}
&=
-\tcl
\bigl(
  3\V{\Ddot{\mathnormal{r}}}\V{\Dot{\vartheta}}+3\Dot r\V{\Ddot{\vartheta}}
  +
  r\V{\dddot{\vartheta}}-r\V{\Dot{\vartheta}}^3
\bigr)
\end{align*}
We seek the motion for $t\gg\tcl$.
The self-force, on the right-hand side, is to be treated as a perturbation
term.
In the lowest order of approximation, with the right-hand side set
to zero, for circular motion the solution is
\[
r=a
~,\quad\V{\Dot{\vartheta}}=\Omega
\]
where $a$ is the radius and $\displaystyle{\Omega=\sqrt{\elec^2/ma^3}}$
is the circular frequency, and $\tcl\Omega\lll1$.

Iterating this solution into the right-hand side gives
\begin{align*}
\V{\Ddot{\mathnormal{r}}}-r\V{\Dot{\vartheta}}^2+a^3\Omega^2\frac{1}{r^2}
&=
0									\\
2\Dot r\V{\Dot{\vartheta}}+r\V{\Ddot{\vartheta}}
&=
a\tcl\Omega^3
\end{align*}
The trial solution $r=a\bigl(1+\alpha t\bigr)$,
$\V{\Dot{\vartheta}}=\Omega\bigl(1+\beta t\bigr)$, with $\alpha$ and
$\beta$ constants, yields
\[
r=a\bigl(1+2\tcl\Omega^2t\bigr)
\quad\text{and}\quad
\Dot{\vartheta}=\Omega\bigl(1-3\tcl\Omega^2t\bigr)
\]
accurate to terms $O(\tcl\Omega)$ in the
interval $\tcl\ll t\ll1/\tcl\Omega^2$.

Following an initial interval of $O(\tcl)$, the motion takes place
in the plane defined by the angular momentum.
Initial circular motion becomes an expanding spiral with decreasing
particle speed and angular velocity.
However, the total energy and the
angular momentum both increase, driven by the Coulomb field.


\section{Conclusion}

Our conceptual shift is to treat the physical particle as a composite
of the (bare) source particle together with the energy-momentum that is
\emph{reversibly} generated in its self-field by its motion.
This energy-momentum reversibly generated in the self-field is bound to
its source particle and is manifested as kinetic energy.
Accordingly, the total kinetic energy of the physical particle
is the sum of the inertial kinetic energy and the bound kinetic
energy of the self-field.
The energy-momentum \emph{irreversibly} generated in the self-field
is propagated from the point at which it is generated as
radiation.

In order to be consistent with the total kinetic energy of the
physical particle being defined as the sum of the kinetic energy
generated in the self-field and the inertial kinetic energy of the
source particle, the corresponding four-forces are summed---see
\eqref{eq: T} and~\eqref{eq: shift}---and this
results in the self-force having its sign opposite to that in the
Lorentz--Dirac equation.
Thus, as a direct consequence of our conceptual shift, we find
Dirac's change of sign for the self-force.

The self-force is the highest order (third) derivative in the equation
of motion and is, in general, a very small perturbation on the other
forces.
The solutions to the applications reveal that the underlying physical
explanation for the behaviour of the particle is that the self-force
extracts energy-momentum from the applied force and converts it into
radiation and kinetic energy for the particle.
It is the latter characteristic of the self-force that causes a particle in
orbital motion to spiral outwards, albeit very slowly.

The new equation of motion, solved as an initial value problem,
has solutions giving `reasonable behaviour' with the physical
interpretation of the self-force and its effects being
straightforwardly comprehensible.
The theory is mathematically well-defined, self-consistent, and
conforms to all well-established principles, including the principle
of causality.

\section*{Acknowledgement}

I am indebted to
Professor C.~W.~Kilmister for his interest and encouragement,
Dr~J.~M. Loveluck for his comments on an early draft, and
Professor G.~L.~Sewell for his generous and detailed comments.

\vspace{-0.5ex}

\bibliography{/Users/mao/Work/Writing/bibliographies/papers,/Users/mao/Work/Writing/bibliographies/books}

\end{document}